\documentclass{aa}  
\usepackage{graphicx}
\usepackage{txfonts}
\usepackage{xcolor}
\begin{document} 

   \title{On the oligarchic growth in a fully interacting system}

   \author{Z. Dencs
          \inst{1}
          \inst{2}
          \and
          Zs. Reg\'aly\inst{1}
          }

   \institute{Konkoly Observatory, Research Centre for Astronomy and Earth Sciences, 1121, Budapest, Konkoly Thege Mikl\'os \'ut 15-17, Hungary\\
              \email{dencs.zoltan@csfk.mta.hu}
         \and
             Department of Astronomy, E\"otv\"os Lor\'and University, 1117, Budapest, Hungary\\
             }

   \date{Received XXX; accepted YYY}
 
  \abstract
   {In the oligarchic growth model, protoplanets develop in the final stage of planet formation via collisions between planetesimals and planetary embryos. The majority of planetesimals are accreted by the embryos, while the remnant planetesimals acquire dynamically excited orbits. The efficiency of the planet formation can be defined by the mass ratio between formed protoplanets and the initial mass of the embryo--planetesimal belt.}
   {In numerical simulations of the oligarchic growth, the gravitational interactions between planetesimals are usually neglected due to computational difficulties. In this way, computations require fewer resources. We investigated the effect of this simplification by modeling the planet formation efficiency in a belt of embryos and self-interacting or non-self-interacting planetesimals.}
   {We used our own developed GPU-based direct N-body integrator for the simulations. We compared 2D models using different initial embryo number, initial planetesimal number, and total initial belt mass. For limited cases, we compared 2D to 3D simulations.}
   {We found that planet formation efficiency is higher if the planetesimal self-interaction is taken into account in models that contain the commonly used 100 embryos. The observed effect can be explained by the damping of planetesimal eccentricities by their self-gravity. The final numbers of protoplanets are independent of planetesimal self-gravity, while the average mass of the formed protoplanets is larger in the self-interacting models. We also found that above 200 embryos, the non-self-interacting and self-interacting models give qualitatively the same results. Our findings show that the higher the initial mass of the embryo--planetesimal belt, the higher the discrepancy is between models that use self-interacting or non-self-interacting planetesimals. The study of 3D models showed quantitatively same results as 2D models for low average inclination. We conclude that below 200 initial embryo number planetesimal self-interaction is important to include in both 2D and 3D models.}


   \maketitle

\section{Introduction}

The formation of terrestrial planets can last for $10^8$ years (see, e.g., \citealp{Safronov1972}). The early phase of the planetary assembly can be described as a runaway growth process, in which the high mass bodies accrete more mass and grow rapidly due to gravitational focusing \citep{Greenbergetal1978}. The building blocks of planets are the km-sized planetesimals, which were formed from the dust component of the protoplanetary disc \citep{KokuboIda2002}. Runaway growth is followed by the next step of final assembly, the oligarchic growth, in which the gravitational focusing is damped, and the velocity dispersion between the neighboring bodies increases due to the perturbation of the larger bodies \citep{KokuboIda1998}. As a result, embryos are formed, which are massive enough to significantly perturb the planetesimals' orbits \citep{IdaMakino1993}. Subsequent collisions of embryos and planetesimals form the precursors of planets called protoplanets \citep{Rafikov2003}.

The mean iteration time of a direct N-body simulation scales with the number of embryos ($N_e$, usually a hundred) and the number of planetesimals ($N_p$, usually a thousand) as $(N_e+N_p)^2$. However, the iteration time can be lowered to $\sim N_e^2+2N_eN_p$ in a non-self-interacting model, where the planetesimals' self-gravity is neglected or approximated.

The final assembly phase is used to be studied with direct N-body methods in a gas-free environment. \citet{Raymondetal2009} investigated the terrestrial planet formation in the inner Solar System after the dissipation of the nebular gas with the direct N-body CPU code, Mercury of \citet{Chambers1999}. In their model, there were 85--90 embryos ($5\times10^{-3}-10^{-1}\,M_{\oplus}$), 1000--2000 planetesimals ($2.5\times 10^{-3}\,M_{\oplus}$) as well as a Jupiter and a Saturn mass giant. Embryo--embryo, and embryo--planetesimal interactions were modeled, while the planetesimal self-interaction was neglected due to calculation difficulties. Using a similar simplification for modelling the terrestrial planet formation with Mercury code, \citet{Roncoetal2015} presented  models that contain 45 embryos ($6\times10^{-2}-4.7\times10^{-1}\,M_{\oplus}$) and 1000 non-self-interacting planetesimals ($2.68\times10^{-3}-2.1\times10^{-2}\,M_{\oplus}$). In a recent study, \citet{LykawkaIto2019} investigated the formation of the terrestrial planets in the Solar System using a modified version of Mercury integrator \citep{HahnMalhotra2005,KaibChambers2016}. \citet{LykawkaIto2019} performed 540 simulations with 20--116 embryos and 2230--7000 non-self-interacting planetesimals.

\citet{LevisonMorbidelli2007} investigated the formation of ice giants and the Kuiper Belt with SyMBA code \citep{LevisonDuncan2000}. They performed models with 3--6 embryos ($1\,M_{\oplus}$) and rings of 1000--32700 planetesimals ($10^{-2}-8\times10^{-2}\,M_{\oplus}$). In their models, planetesimal self-interaction was approximated by a radial force component based on the formalism of \citet{BinneyTremaine1987}. \citet{LevisonMorbidelli2007} found that neglecting self-gravity resulted in unphysical embryo motion because planetesimals were trapped in 1:1 mean-motion resonance with embryos. This phenomenon is removed by approximating planetesimal self-gravity.

\citet{Quintanaetal2007} investigated the terrestrial planet formation in a binary star system. Their model contained a relatively low number of particles, 14 embryos ($9.3\times10^{-2}\,M_{\oplus}$) and 140 self-interacting planetesimals ($9.3\times10^{-3}\,M_{\oplus}$) assuming an 50/50 embryo-to-planetesimal mass ratio. In another fully interacting model, \citet{Morishimaetal2008} simulated the formation of terrestrial planets. To model the very first phase (runaway growth), an approximation of direct force calculation of 1000--5000 self-interacting planetesimals was applied (parallel tree code of \citealp{Richardsonetal2000}). \citet{FanBatygin2017} investigated the effect of planetesimal self-gravity on the evolution of giant planets' and Kuiper Belt objects' orbits. They used QYMSYM GPU-based N-body code \citep{MooreQuillen2011} and compared self-interacting to non-self-interacting models, which contain 1000 planetesimals and the four giant planets. The orbital architecture of the final systems was similar, however, instabilities led to the destruction of the system in more cases with self-interacting models than with non-self-interacting ones.

In this study, we investigate how planetesimal self-interaction affects the planet formation efficiency (mass transfer efficiency between the initial bodies and the protoplanets) in the final assembly phase of planet formation. For the direct N-body calculations we use Graphics Processing Units (GPU) which enables us to integrate fully interacting systems. We present models containing embryos embedded in disc of self-interacting or non-self-interacting planetesimals. The effect of the initial embryo-to-planetesimal mass ratio is also investigated in both cases. Models with different embryo and planetesimal numbers are compared. The outline of this letter is the following. In Section\,2, we present the applied numerical integrator and the initial conditions. Section\,3 deals with the results of simulations and the discussion. In Section\,4, we summarize our main conclusions.

\section{Numerical simulations}

We modeled the oligarchic growth using N-body simulations. Initially, three types of bodies are presented in the system: a star, planetary embryos, and planetesimals. Thus, the simulated system does not contain gas. This model resembles the environment of the last phase of planet formation. 

In this investigation, two distinct models were compared in which the system contains: gravitationally self-interacting planetesimals (referred to as INT); and non-self-interacting planetesimals (NON). The gravitational interaction between planetesimals is neglected, while the embryo--embryo, embryo--star, embryo--planetesimal, star--planetesimal interactions are taken into account in NON models. In INT models all bodies are in interaction with each other. 

We used our own developed numerical integrator for the calculations, HIPERION\footnote{https://www.konkoly.hu/staff/regaly/research/hiperion.html}, which is a GPU-based direct N-body integrator. We applied a 4$^{\mathrm{th}}$ order Hermite scheme for all simulations \citep{MakinoAarseth1992}. The calculations were performed on NVidia Tesla K20 and K40 GPU cards. At each integration step the optimal time-step is given by 4$^{\mathrm{th}}$ order version of the generalized Aarseth scheme \citep{PressSpergel1988,MakinoAarseth1992}. We verified the $\eta$ parameter that controls the precision of the integrator in the range of $0.0025 \leq \eta \leq 0.04$. We found that the evolution of the total embryo mass was independent of $\eta$. The precision of the integrator can be characterized by the relative energy error of the system, which is given by the ratio of instantaneous and initial system’s total (potential plus kinetic) energy \citep{NitadoriMakino2008}. By using a canonical setting $\eta=0.02$ the relative energy error is found to be in the range of $10^{-6}-10^{-3}$. 

In our simulations the close encounter of different bodies can result in the following phenomena: bodies collide and merge, engulfed by the star, or scattered out from the system. Only embryo--planetesimal or embryo--embryo collisions were allowed if the distance between the center of the two colliding bodies is smaller than their mutual radii. The collisions are assumed to be fully inelastic, i.e. the total mass and momenta of colliding bodies are conserved. The total energy of the system is corrected by the energy loss that occurred during the collisions. The density of the bodies is assumed to be constant, therefore the radius of the forming body is determined by the added mass. Thus, the fragmentation of the bodies is ignored, which has been proved to be a plausible approximation \citep{WetherillStewart1993}. Bodies are removed from the system if their distance to the star is less than 0.1\,au or larger than 10\,au.

The initial structure of the systems is the same both in NON and INT simulations. The host star is represented by a $1\,M_{\odot}$ central body. The inner and the outer boundary of the simulated embryo--planetesimal belt are 0.4\,au, and 4.2\,au, respectively. The number density of embryos and planetesimals follows a power law of $-1$ as a function of radial distance.

To investigate the statistical properties of the evolving systems first, we ran 10 INT and 10 NON models, each containing a belt of 100 embryos and 1000 planetesimals randomly distributed. The initial position of the embryos and planetesimals was 10 times redistributed. The 10 INT and the 10 NON simulations were modeled in 2-dimensions (2D), where eccentricity and inclination of the embryos, as well as the planetesimals, were set to zero. Additionally, we ran 3-dimensional (3D) models, in which cases the mean eccentricity and inclination of embryos, as well as planetesimals, are in the ranges of $10^{-5}-10^{-1}$ and $5\times10^{-6}-5\times10^{-2}$, respectively.
 
The total mass of the initial embryo--planetesimal belt is based on a scaled version of the Minimum Mass Solar Nebula \citep{Weidenschilling1977,Hayashi1981} model. The total mass of the embryo--planetesimal belt is 7.35\,M$_{\oplus}$, which is 3 times greater than the dust mass of Solar Nebula in this region, which is a conservative calculation based on \citet{Lissauer1987}. The initial embryo-to-planetesimal mass ratio is 50/50. Embryos and planetesimals have uniform masses, thus the individual initial masses of an embryo and a planetesimal are $4\times10^{-2}\,M_\oplus$, and $4\times10^{-3}\,M_\oplus$, respectively. The radii of the embryos and planetesimals are calculated based on their masses by assuming a uniform 2 gcm$^{-3}$ density, which gives 3000 and 1400\,km in our standard models, respectively.

To understand the effect of planetesimal self-interaction, 2 INT and 2 NON models were applied, in which the initial embryo-to-planetesimal mass ratio was 25/75, and 75/25. The individual masses were corrected according to the mass ratios in these models. 

To investigate the effect of the initial number of embryos and planetesimals we perform additional simulations: 1) 4 INT and 4 NON models consisting of 100 embryos, and planetesimals in the range of 250--4000; 2) 4 INT and 4 NON models with embryo number in the range of 25--400, and 1000 planetesimals. The embryo-to-planetesimal mass ratio is fixed to 50/50.

\section{Results and discussion}

As a result of embryo--embryo and embryo--planetesimal collisions the mass of embryos increases with time. The total mass confined by accreting embryos grows exponentially with time in the beginning. Later the growth rate slows down and the total embryo mass saturates. This phase determines the time–scale of the simulations, which is found to be $2\times10^5$\,years in all 2D models, and $\sim2\times10^{6}$\,years in 3D models in agreement with the results of \citet{KokuboIda2000}. The residual embryos can be considered as protoplanets. 

\subsection{Properties of the final systems}

\begin{figure}
    \begin{center}                                                         
        \includegraphics[width=\hsize]{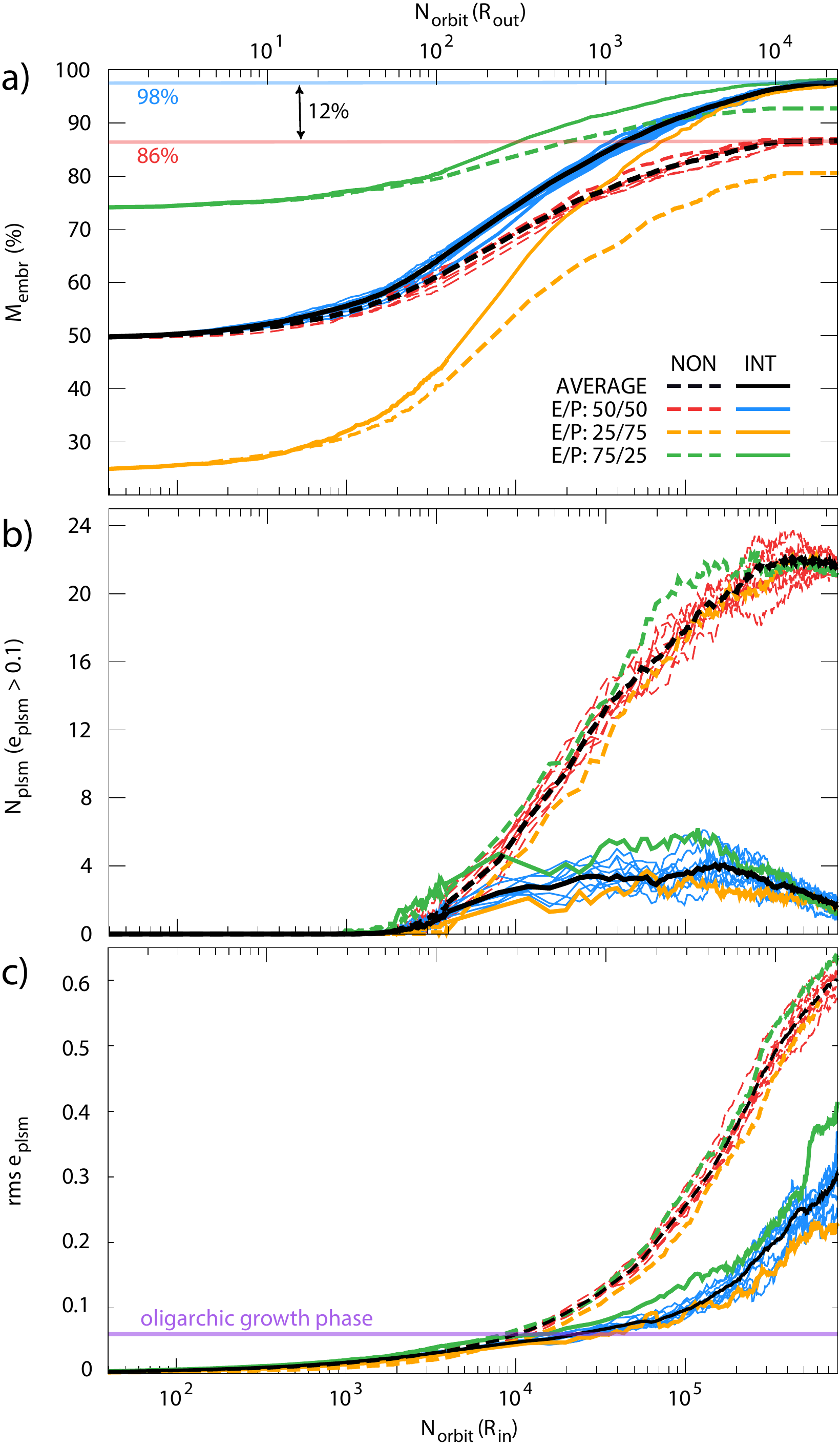}
    \end{center}
    \caption{Panel\,a) shows the evolution of the total mass confined by accreting embryos. The mass is normalized by the initial mass of the embryo--planetesimal belt. The top and bottom horizontal axes display the number of orbits at the inner and the outer edge of the belt, respectively. Red dashed and blue solid lines show the mass evolution of the 50/50 mass ratio in NON and INT models, respectively. Black lines show the average for 10--10 models in both cases. Orange and green lines indicate models where 25/75, and 75/25 initial embryo-to-planetesimal mass ratio is assumed. Panel\,b) shows the evolution of the number of HEP population normalized with the total number of planetesimals in all models. Panel\,c) displays the evolution of the planetesimal rms eccentricity. Above the horizontal purple line, the system is in the oligarchic growth phase.}
    \label{fig:stat}
\end{figure}

\begin{figure}
    \begin{center}                                                         
        \includegraphics[width=\hsize]{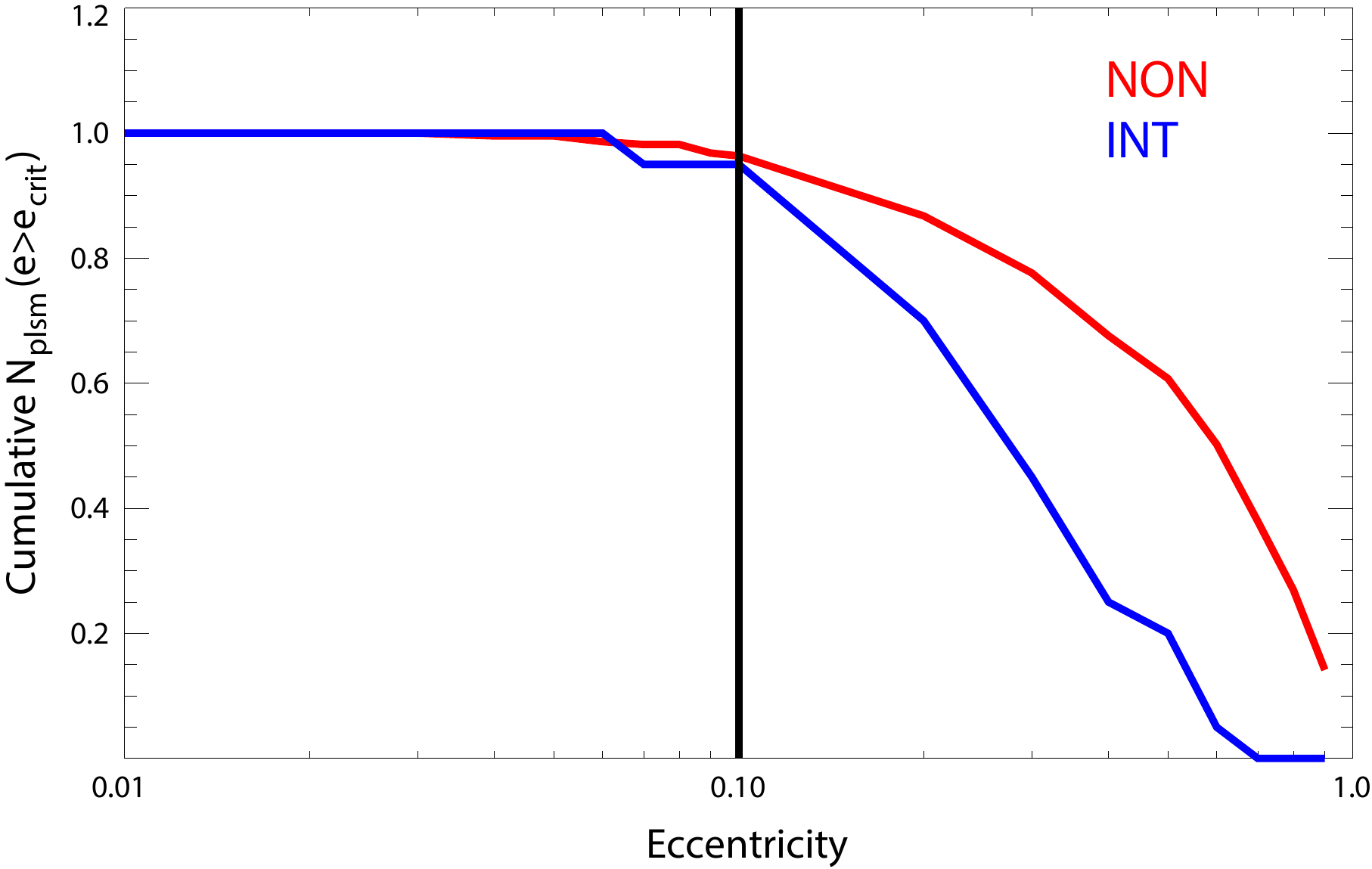}
    \end{center}
    \caption{The number of planetesimals (normalized with the initial number of planetesimals) having an average eccentricity larger than a given eccentricity limit. Red and blue curves show the number of planetesimals in NON and INT models respectively (we note that the NON curve was divided by 10 for the comparison with the INT curve). The vertical black line indicates the critical eccentricity value used to determine the HEP number.}
    \label{fig:crit}
\end{figure}

First, we analyze 10--10 2D models, where 100 embryos and 1000 planetesimals were in the system initially. In INT models nearly all the mass is converted to protoplanets. In NON models, the number of the remnant planetesimals is significant at the end of simulations. As a result, the total mass of protoplanets is lower in NON than in INT models, as can be seen in panel\,a) of Fig.\,\ref{fig:stat}. The total protoplanetary mass (calculated for the 10--10 distinct models, and represented by black lines in panel\,a) of Fig.\,\ref{fig:stat}), is about 12\,percent higher in INT than in NON models.

Regarding the number of formed protoplanets, we found an average of 11.6 in INT and 13.1 in NON models. The difference in the number of remnant planetesimals in the two models is significant: 18.5 in INT, while 237.4 in NON models. Namely, nearly 13 times more planetesimals are accreted by protoplanets in INT than in NON models. As a consequence, the average mass of the protoplanets is $0.67\,M_{\oplus}$ and $0.53\,M_{\oplus}$ in INT and NON models, respectively. The mean separation of the protoplanets’ orbit is 0.54\,au and 0.35\,au in INT and NON models, respectively. The mean eccentricity of the protoplanets is 0.12 and 0.1 in INT and NON models, respectively.

We also investigated the effect of the initial embryo-to-planetesimal mass ratio (see 25/75 and 75/25 mass ratios with orange and green lines in Fig.\,\ref{fig:stat}). We found that the difference between INT and NON models in the total mass confined by protoplanets increases with the decreasing initial embryo mass. We must Emphasize that in INT models the total mass transformed to protoplanets ($\sim$98\,percent) does not depend on the initial embryo mass. 

Now, we discuss our findings regarding the efficiency of mass conversion to protoplanets. As a result of embryo--planetesimal interactions, the eccentricity of planetesimals is excited to high values and a population of highly eccentric planetesimals (HEP) forms. We found that the cumulative number of planetesimals with a given eccentricity is constant up to a critical value, then steeply drops (see Fig.\,\ref{fig:crit}). The value of the critical eccentricity is found to be 0.1 for all models. Panel\,b) of Fig.\,\ref{fig:stat} shows the evolution of the number of HEP. It is clearly seen that the growth rate in the HEP number is much higher (causing 10 times higher HEP number) in NON than in INT models. Simulations assuming different embryo-to-planetesimal mass ratios show the same effect. The mean eccentricity of planetesimals is 0.09 and 0.44 at its maximum in INT and NON models, respectively.

The difference in the total mass of the HEP population in INT and NON models explains the difference in the total mass of protoplanets formed in INT and NON models. In our simulations, HEPs are accreted with a lower probability, which explains the difference in the total mass of protoplanets in INT and NON models. 

In the oligarchic growth phase, the velocity dispersion, as well as the eccentricity of the planetesimals increase due to the excitation of the embryos. Panel\,c) of Fig.\,\ref{fig:stat} shows the evolution of the root mean square (rms) eccentricity of the non-accreted planetesimal population for the 10 INT, 10 NON standard models, and the 25/75, as well as the 75/25 mass ratio models. One can see in panel\,c) that the rms eccentricity increases with time, while the number of the remnant planetesimals decreases. As a result, the eccentricity of the relatively few non-accreted planetesimals is high. We note that highly eccentric planetesimals will not be accreted by the embryos in the later phase.

One can also see in panel\,c) of Fig.\,\ref{fig:stat} that the rms eccentricity of the planetesimals grows slower in INT than in NON models due to the self-gravity of planetesimals. On average, the rms eccentricity (which is not saturated) is about twice as much in NON than in INT models. The start of the oligarchic growth phase is indicated by a purple horizontal line in panel\,c). It can be seen that all models are in the oligarchic growth phase for most of the integration time: the orderly growth phase lasts only for $\sim 10^4$ orbits, while the oligarchic phase lasts for $\sim 7\times10^5$ orbits at the inner edge of the embryo--planetesimal belt.

\subsection{Effect of the initial embryo and planetesimal number}

\begin{figure}[!h]
	\begin{center}  
    	\includegraphics[width=\hsize]{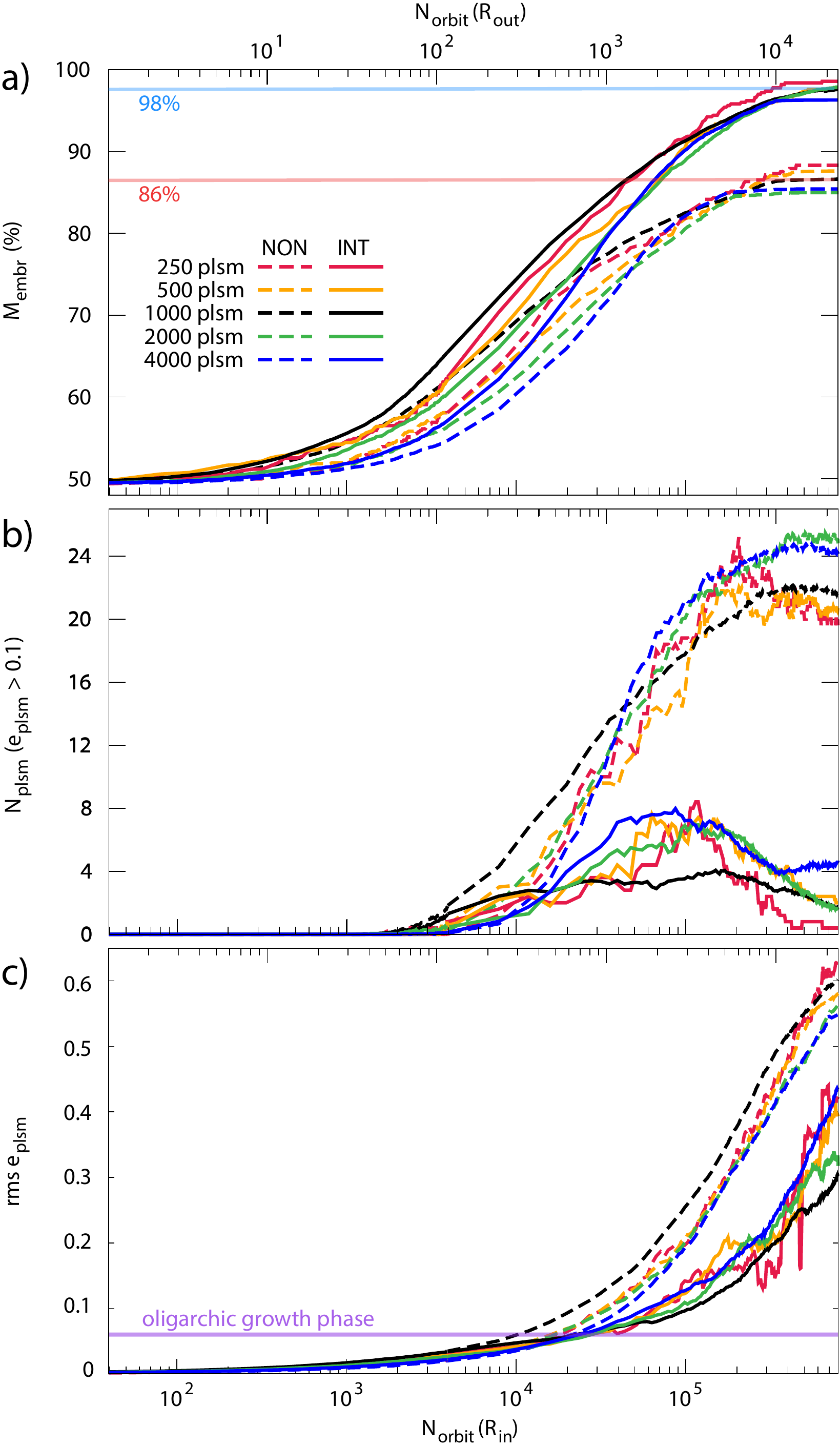}
    \end{center}
	\caption{Same as Fig.\,\ref{fig:stat} but models use different initial numbers of planetesimals while the number of embryos is fixed to 100. Colors represent models with different initial numbers of planetesimals in the rage of 250--4000.}
	\label{fig:plsm}
\end{figure}

\begin{figure}[!h]
	\begin{center}  
    	\includegraphics[width=\hsize]{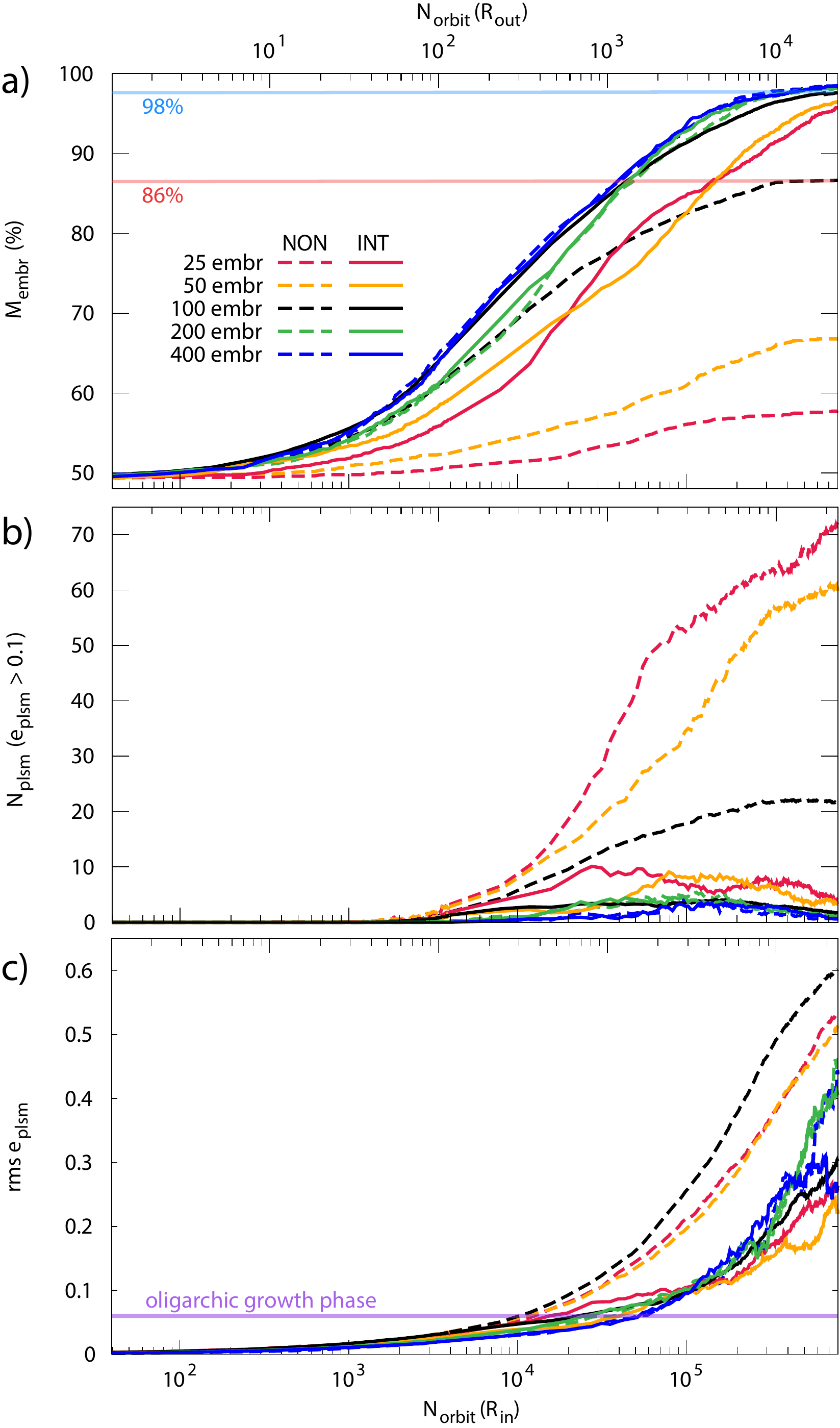}
    \end{center}
	\caption{Same as Fig.\,\ref{fig:stat} but models use different initial numbers of embryos while the number of planetesimals is fixed to 1000. Colors represent models with different initial numbers of embryos in the range of 25--400.}
	\label{fig:embr}
\end{figure}

\begin{figure*}
    \begin{center}                                                         
        \includegraphics[width=\hsize]{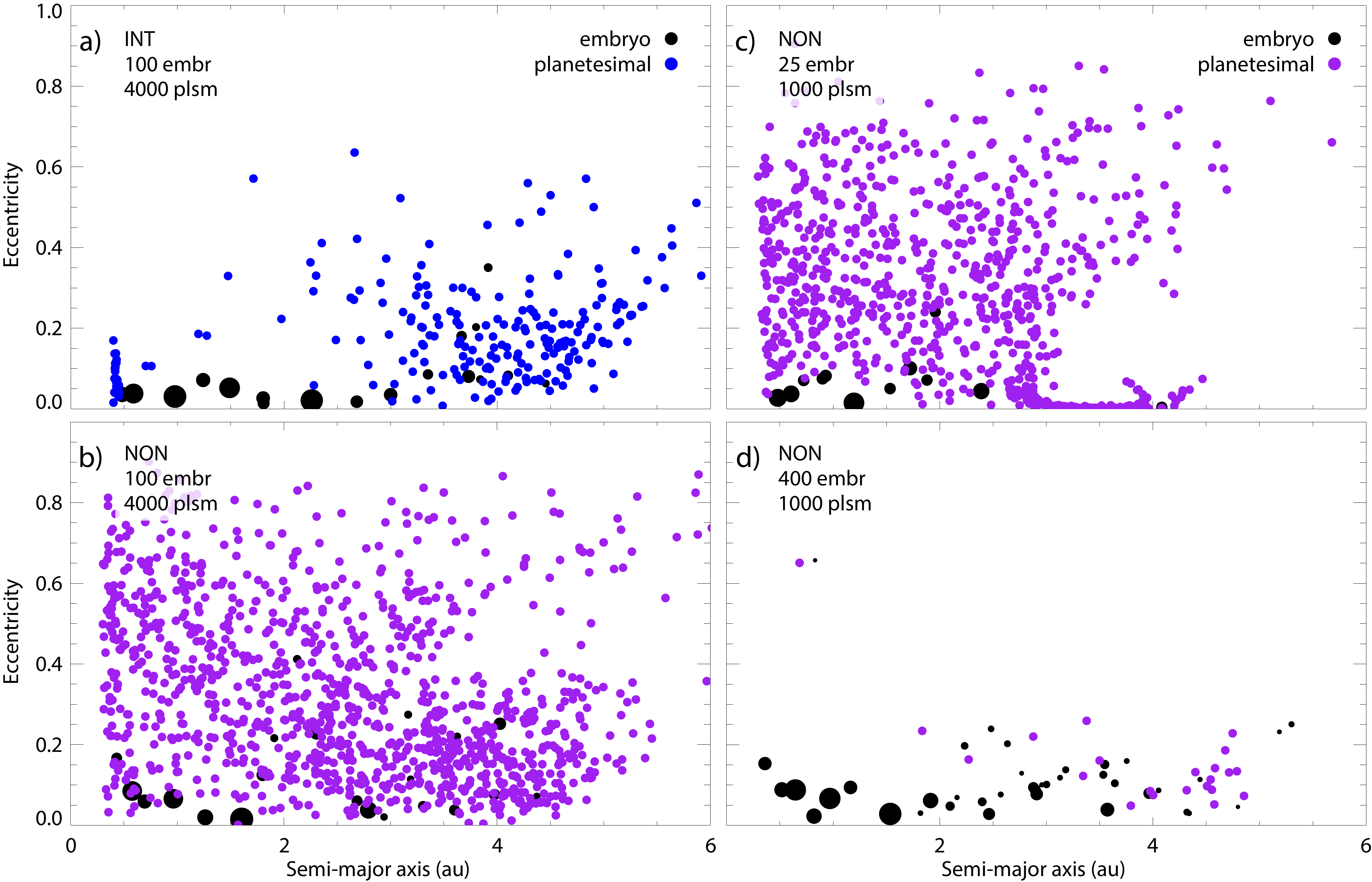}
    \end{center}
    \caption{The eccentricity of embryos and planetesimals as a function of the semi-major axis at the $3.1 \times 10^{5{\mathrm{th}}}$ orbit at the inner edge. Panel\,a) and b) show an INT and a NON model, respectively, initially with 50/50 mass ratio, 100 embryos (black dots) and 4000 planetesimals (blue dots in the INT, purple dots in the NON model). Panel\,c) and d) show NON models initially with 25 and 400 embryos, respectively, as well as with 1000 planetesimals. The size of black dots is proportional with the $1/3^{\mathrm{th}}$ power of the embryo mass.}
    \label{fig:ae}
\end{figure*}

In the following, we investigate the effect of the number of bodies in the system. Panel\,a) of Fig.\,\ref{fig:plsm} displays the evolution of the total embryo mass in models where the initial planetesimal number changes in the range of 250--4000, while the embryo number is the same (100) as in previous models. It can be seen that the total protoplanetary mass only depends weakly on the planetesimal number. Thus, the departure of INT and NON models is independent of the initial planetesimal number. Regarding the number of HEP, we have found that the eccentricity evolution of planetesimals is also weakly dependent on the planetesimal number, see panel\,b) of Fig.\,\ref{fig:plsm}. We have not found any correlation between the number of initial and remnant planetesimals either in NON or INT models. Based on the rms eccentricity evolution of planetesimals, NON models require only half the time to enter the oligarchic growth phase compared to INT models, see panel\,c) of Fig.\,\ref{fig:plsm}. This phenomenon is independent of the initial number of planetesimals.

Contrary to previous findings, the initial number of embryos affects the results in NON models, see Fig.\,\ref{fig:embr}. In NON models, where the initial embryo number changes in the range of 25--400 while the planetesimal number is 1000, we observe strong dependence of the total protoplanet mass on the initial embryo number. The larger the embryo number, the larger the mass transported to protoplanets is. Thus, in NON simulations the planet formation efficiency is increasing with the initial embryo number. However, above a critical embryo number (found to be 200) the mass conversion is $\sim98$\,percent (panel\,a) of Fig.\,\ref{fig:embr}), which means that INT and NON models give the same results. One can see in panel\,b) of Fig.\,\ref{fig:embr} that the number of HEP decreases with the increasing initial number of embryos and above 200 practically no high eccentric planetesimals can be found. In these cases, the total mass of the formed protoplanets is equal in INT and NON models. According to panel\,c) of Fig.\,\ref{fig:embr} the evolution of planetesimal rms eccentricity deviates in NON and INT models. However, above 200 initial embryos, both models show the same rms eccentricity evolution for the majority of the simulation.

In NON models, we have found that the lower the initial number of embryos, the higher the planetesimal eccentricity is. We note that in our models, the individual embryo mass is inversely proportional to the initial number of embryos because the embryo--planetesimal belt mass is kept fixed. Moreover, the high eccentricity excitation prevents the frequent accretion of the planetesimals in the assembly phase. As a result, the increasing embryo number leads to a weak excitation, therefore, the difference in the accretion efficiency of planetesimals between INT and NON models decreases, too. Our results also show that the number of HEP is very low in all INT models independently of the initial number of embryos. This can be explained by the eccentricity damping effect of planetesimal self-gravity. 

In INT models the results are independent of the initial embryo mass. This is due to the fact that planetesimal self-gravity strongly damps their eccentricity. Therefore, the individual embryo mass has no effect on the mass conversion to protoplanets.

The damping effect of planetesimal self-gravity is shown in Fig.\,\ref{fig:ae}, which displays the semi-major axis--eccentricity diagram of embryos and planetesimals in an INT (panel\,a) and three NON (panel\,b--d) models at the same evolutionary phase. Comparing panel\,a) and b), one can see that about 3/4 and 1/20 of planetesimals have been accreted in INT and NON models, respectively. The mean eccentricity of non-accreted planetesimals is 0.19 and 0.34 in INT and NON models, respectively. This is explained by the fact that the low-eccentric planetesimals are accreted by the embryos with higher probability than the high-eccentric planetesimals. 

To demonstrate the eccentricity excitation effect of the individual embryo mass, panel\,c) and d) of Fig.\,\ref{fig:ae} show a low and a high initial embryo number model, respectively. One can see that a large number of non-accreted planetesimals are in the system in case of low initial embryo number (high individual embryo mass), see panel\,c). Contrarily, for high initial embryo number, practically no non-accreted planetesimals are present, see panel\,d). We emphasize that the INT model shows similar conversion efficiency as the high initial embryo number NON model, see panel\,a) and d).    

\subsection{Effect of the initial mass of embryo--planetesimal belts}

The effect of the initial mass of embryo--planetesimal belts on the evolution of the total protoplanetary mass in INT and NON models was also investigated. We performed simulations with a 22.05\,M$_{\oplus}$ initial mass belt being three times more massive than what was used in previously presented models. In these models, 25--400 initial embryo number, a constant 1000 planetesimals, and a 50/50 embryo-to-planetesimal mass ratio were used. 

\begin{figure}
	\begin{center}                                                         
    	\includegraphics[width=\hsize]{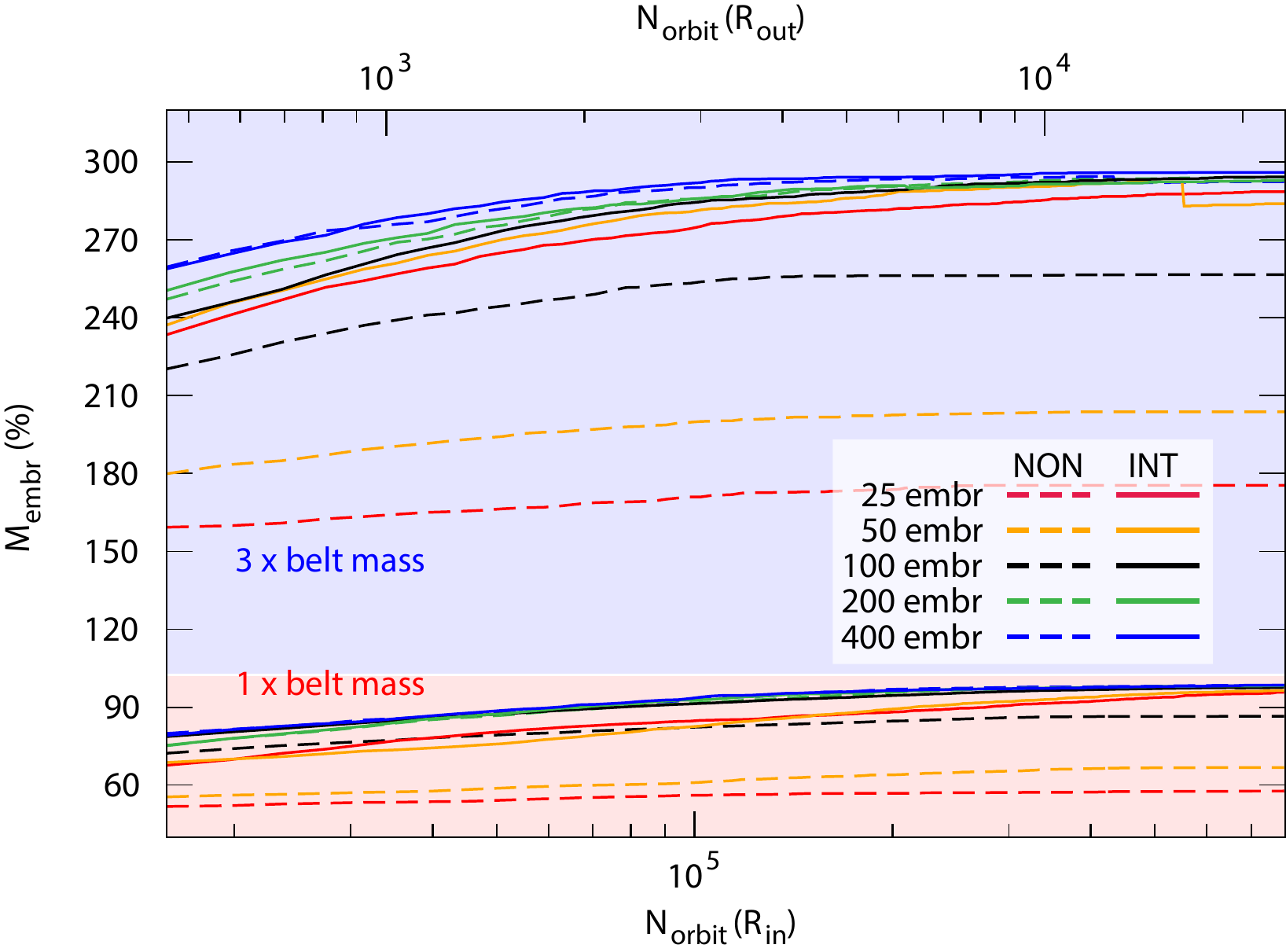}
    \end{center}
	\caption{The evolution of the total mass confined by the embryos in INT and NON models. Red and blue shaded regions indicate the previously used embryo--planetesimal belt mass (7.35\,M$_{\oplus}$) and the three times higher belt mass (22.05\,M$_{\oplus}$), respectively. Colored curves represent models with different initial numbers of embryos (indicated in the key) in the range of 25--400 with a constant 1000 planetesimals.}
	\label{fig:mass}
\end{figure}

Figure\,\ref{fig:mass} shows the evolution of the protoplanet mass in high mass and standard mass models. One can see that the departure of INT and NON models in the total protoplanetary mass is higher in the more massive, than in the less massive belts as long as less than 200 initial embryos are used. We have also found that INT and NON models show the same results for more than 200 initial embryos independently of the initial belt mass.

\subsection{2D and 3D embryo--planetesimal belts}

The time required to saturate the total protoplanetary mass strongly depends on the average orbital inclination of the interacting bodies: the higher the average inclination, the slower the saturation is. This is because the probability of collisions is lower if the mutual inclination of embryos and planetesimals is higher. Considering this fact, simulations presented so far were calculated in 2D. In this case, the orbits of interacting bodies are coplanar. To investigate the effect of planetesimal self-gravity in 3D, we also investigate 5 additional models that start with dynamically excited embryo--planetesimal belts. The eccentricities and the inclinations follow Rayleigh distribution \citep{Lissauer1993} assuming that $\langle e^2 \rangle^{1/2} \simeq 2\langle i^2 \rangle^{1/2}$ of \citet{IdaMakino1993}. We investigated 5 different models where $10^{-5} \leq \sigma_e \leq 10^{-1}$ and $5\times 10^{-6} \leq \sigma_i \leq 5\times 10^{-2}$, the initial mass ratio is 50/50, and the initial belt consists of 100 embryos as well as 1000 planetesimals. 

\begin{figure}[!h]
	\begin{center}  
    	\includegraphics[width=\hsize]{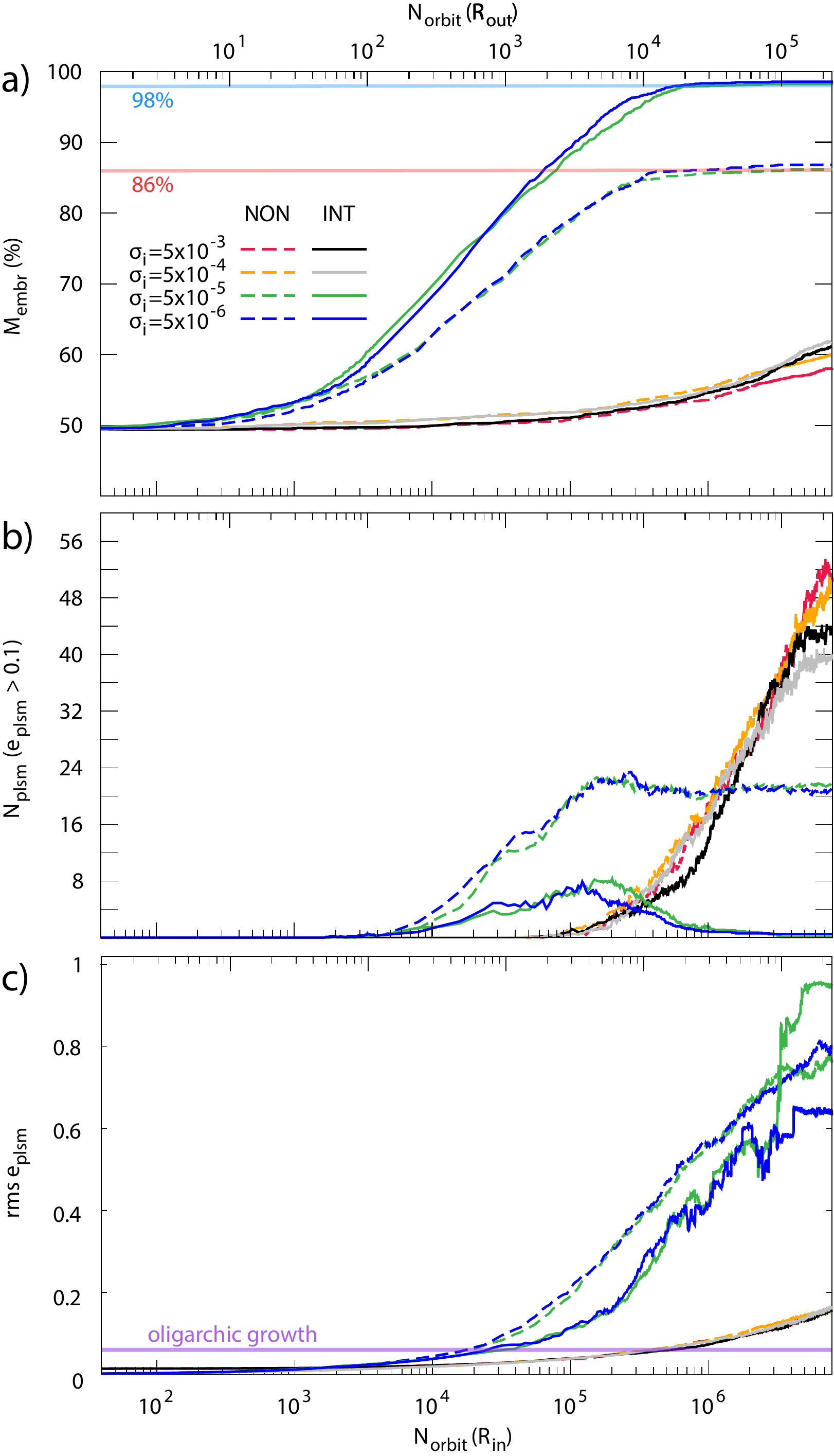}
    \end{center}
	\caption{Same as Fig.\,\ref{fig:stat} but here 3D models are presented. Colors represent embryo--planetesimal disks with different initial mean eccentricity and inclination in INT and NON models with initially 100 embryos, and 1000 planetesimals. The embryo-to-planetesimal mass ratio is 50/50.}
	\label{fig:inc}
\end{figure}

Panel\,a) of Fig.\,\ref{fig:inc} shows the evolution of total mass confined by the accreting embryos in INT and NON 3D models. For the initially moderately excited belts ($\sigma_i \leq 5\times10^{-5}$) the total protoplanetary masses are saturated and show different conversion efficiency (same as in 2D models, 12\,\%) in INT and NON models. We note that the time-scale of the embryo growth is about 2 times slower in 3D models. The growth rate decreases by $\sigma_i$. Therefore, models with $\sigma_i \geq 5\times10^{-4}$ are not saturated.

Panel\,b) of Fig.\,\ref{fig:inc} shows the evolution of the HEP number, which begins to grow at $10^{4{\mathrm{th}}}$ orbit in $\sigma_i \leq 5\times10^{-5}$ models, while it begins to grow at later times, $10^{5{\mathrm{th}}}$ orbit at the inner edge of the embryo--planetesimal belt in $\sigma_i \geq 5\times10^{-4}$ models. It is also visible that the number of HEP is at least two times higher in highly excited models than the peak of HEP number in moderately excited models at the end of the simulations. The difference between INT and NON 3D models can be explained by the same phenomenon (eccentricity damping of self-gravity) which was identified in 2D models. 

Panel\,c) of Fig.\,\ref{fig:inc} displays the evolution of the planetesimal rms eccentricity. One can see that the $\sigma \leq 5\times10^{-5}$ models saturate at $\sim 4\times10^6$ orbit (at the inner edge of the belt), while the highly excited models do not saturate in the investigated time-scale. In moderately excited models the oligarchic phase occurs about two times later compared to 2D models. To enter the oligarchic growth phase at least 10 times more number of orbits are required for the initially highly excited models than for the moderately excited models (see panel\,c) of Fig.\,\ref{fig:inc}). Note also that, the rms eccentricity grows faster in moderately excited NON than in INT models.

To summarise our findings, the higher the initial mean inclination of embryo and planetesimal orbits, the slower the growth rate of the embryos due to the lower probability of planetesimal accretion. However, we emphasize that neglecting planetesimal self-gravity has the same effect on the embryo growth in 3D as in 2D. 

\section{Conclusions}

In this paper, we investigated the effect of the planetesimals’ self-gravity on the final assembly phase of planet formation. In these models, the bodies of embryo and planetesimal populations are collided with each other to form growing embryos and finally, protoplanets. Two types of models are compared: 1) self-interacting planetesimals (INT), and 2) non-self-interacting planetesimals (NON). The circumstellar embryo--planetesimal belt consists of bodies with uniform mass. The total mass of the embryo--planetesimal belt is based on the MMSN model \citep{Weidenschilling1977,Hayashi1981}. We ran models with and without planetesimal self-gravity. We investigated the effect of both the initial number of planetesimals (250--4000) and embryos (25--400). The simulations ran $2\times10^5$ years which corresponds to $7.91\times10^5$ and $2.32\times10^4$ orbits at the inner and outer edge of the embryo--planetesimal belt, respectively. Beyond this time only protoplanet collisions are observed and the planetesimal accretion is absent. We ran additional models with three times higher mass embryo--planetesimal belt as well as dynamically excited belts. We investigated the evolution of the planetesimal rms eccentricity. Based on our investigation, the modeled systems are dominantly in the oligarchic growth phase during the simulations. Our main results are the following. 

\begin{enumerate}
\item Simulations started with a relatively large number of embryos (100) and planetesimals (1000) showing that the total mass of the protoplanets is about 12\,percent higher in INT than in NON models. The number of remnant planetesimals (which are not accreted) is about an order of magnitude higher in NON than in INT simulations.

\item Low-eccentric planetesimals are easily accreted by the embryos, while the high eccentric ($e\gtrsim0.1$) planetesimals remain in the system for a long time. Thus, the total protoplanetary mass at the end of the final assembly phase depends on the strength of the eccentricity excitation of planetesimals. The latter is highly damped in INT models, which results in 98\,percent efficiency in mass conversion from embryo and planetesimals to protoplanets. This can be explained by the effect of the eccentricity damping of planetesimal self-gravity.

\item The total mass of the protoplanets is independent of the initial number of planetesimals (in the range of 250--4000) both in INT and in NON models that contain 100 embryos.

\item In the simpler NON models, the total mass of protoplanets is proportional to the initial embryo number (with a constant planetesimal number). However, above a critical embryo number (200), INT and NON models present the same results.

\item The higher the initial mass of the embryo--planetesimal belt, the larger the discrepancy is between INT and NON models in terms of the final mass confined by protoplanets, assuming a commonly used initial embryo ($\leq 100$) and planetesimal (1000) number.

\item 3D simulations give qualitatively the same results as 2D models. For low average inclination 3D models we find quantitative equivalence with 2D results.
\end{enumerate}

We emphasize that terrestrial planet formation is not finished at the end of our simulations. Protoplanets still perturb each others’ orbit and may collide with each other leading to the formation of planets. Therefore, the architectural differences found in INT and NON models (0.54 and 0.35\,au mean separation; $0.67$ and $0.53\,M_{\oplus}$ mean mass; 0.12 and 0.1 mean eccentricity of protoplanets) can not be interpreted as appropriate orbital elements of the final planetary system. Terrestrial planet formation may cover a very long time, up to $10^8$ years, however, we investigated a shorter time-scale by three orders of magnitude. Thus, to reveal the real architectural difference in INT and NON models requires further study, where the protoplanetary systems are integrated further in time.

Here we mention some caveats of our models. First, we ignored the effect of the fragmentation during collisions which may affect the dynamics of the system and the embryo growth. However, according to \cite{WetherillStewart1993} fragmentation has no significant effect on embryo growth. Secondly, highly excited 3D models were not followed to protoplanetary mass saturation due to computational difficulties. However, based on our moderately excited models we think that our findings can be generalized for 3D cases.

According to our simulations, the mean iteration time (MIT) increases linearly with the number of bodies both in NON and INT models. We emphasize that the slope of the MIT-particle number curve is 1.23 times steeper in INT than in NON models. However, the linearity breaks above 6000 bodies in INT models due to the computational capability of the GPUs used for this investigation. Fortunately, our INT simulations are numerically convergent at a lower number (100 embryo and 1000 planetesimals) of bodies than this limit. As a result, our fully GPU-based direct N-body integrator is an efficient tool for investigating the final assembly of terrestrial planet formation.

We emphasize that the evolution of the protoplanet mass does not depend on the accuracy of the integrator (relative energy error is found to be in the range of $10^{-6}-10^{-3}$), but on the initial number of bodies. We have shown that the initial number of bodies in the N-body models with non-self-interacting planetesimals may have a significant effect on the efficiency of mass transfer from small bodies to protoplanets. We conclude that numerical simulations of the final assembly phase of terrestrial planet formation require a relatively large number of embryos ($\gtrsim200$) if the planetesimal self-interactions are neglected. However, N-body simulations that are not simplified by neglecting planetesimal--planetesimal interactions are numerically convergent with a lower number of embryos. Thus, if the applied hardware for the numerical solution is highly sensitive to the number of particles (e.g., CPU-based direct integrator), we suggest using at least twice as many fully interacting embryos than conventionally applied previously.

\begin{acknowledgements}
We thank the anonymous referee for useful suggestions that helped to improve the quality of the paper. This work was supported by the Hungarian Grant K119993. We acknowledge the Hungarian \emph{National Information Infrastructure Development Program} (NIIF) for awarding us access to the computational resource based in Hungary at Debrecen.
\end{acknowledgements}


\end{document}